\documentclass[twocolumn]{autart}    % Enable this line and disable the
\usepackage{graphicx}          % Include this line if your
\usepackage{epstopdf}
\usepackage{amsmath,amssymb,amsfonts}
\usepackage{verbatim}
\usepackage{enumerate}
\usepackage{bm}
\usepackage{color}
\usepackage{cite}
\usepackage{tikz,times}
\usepackage[colorlinks, linkcolor=blue,
            anchorcolor=blue,
            citecolor=blue,
            urlcolor=blue
]{hyperref}

\theoremstyle{plain}

\newtheorem{lemma}{\textbf{Lemma}}

\theoremstyle{definition}
\newtheorem{definition}{\textbf{Definition}}

\theoremstyle{remark}
\newtheorem{remark}{Remark}

\begin{document}

\begin{frontmatter}

\title{Stabilizing Preparation of Quantum Gaussian States via Continuous Measurement\thanksref{footnoteinfo}} % Title, preferably not more

\thanks[footnoteinfo]{The material in this paper was not presented at any conference.}

\author[a,b]{Liying Bao}\ead{baoliying@amss.ac.cn},    % Add the
\author[a,b,cor1]{Bo Qi}\ead{qibo@amss.ac.cn},               % e-mail address
\author[c]{Daoyi Dong}\ead{daoyidong@gmail.com}  % (ead) as shown
\corauth[cor1]{Corresponding author.}

\address[a]{Key Laboratory of Systems and Control, Academy of Mathematics and Systems Science,\\ Chinese Academy of Sciences, Beijing 100190, China}  % Please supply
\address[b]{University of Chinese Academy of Sciences, Beijing 100049, China}             % full addresses
\address[c]{School of Engineering and Information Technology, University of New South Wales, Canberra ACT 2600, Australia}        % here.

                             % keyword list or with the
                                          % help of the Automatica
                                          % keyword wizard

\begin{abstract}                          % Abstract of not more than 200 words.
This paper provides a stabilizing preparation method for quantum Gaussian states by utilizing continuous measurement. The stochastic evolution of the open quantum system is described in terms of the quantum stochastic master equation.  We present  necessary and sufficient conditions for the system to have a unique stabilizing steady Gaussian state. The conditions are much weaker than those existing results presented in the approach of preparing Gaussian states through environment engineering. Parametric conditions of how to prepare an arbitrary pure Gaussian state are provided. This approach provides more degrees of freedom to choose the system Hamiltonian and the system-environment coupling operators, as compared with the case where dissipation-induced approach is employed. The stabilizing conditions for the case of imperfect measurement efficiency are also presented. These results may benefit practical experimental implementation in preparing quantum Gaussian states.
\end{abstract}

\end{frontmatter}

\section{Introduction}
Continuous variable systems, which are quantum systems with infinite dimensional Hilbert spaces, have been an important platform for quantum cryptography, quantum information and quantum computation \cite{Edwards2005,Handel2005,James2008,Wiseman2010,Nurdin2012,PanZhang2016,MaWoolley2018,GaoZhang2020,GaoPetersen2020,Ghalaii12020,ZhangPan2020,ZhangPetersen2020,GaoDong2021,GuoPeng2021}. Gaussian states, which include a wide and important class of quantum states such as vacuum states, squeezed  vacuum, squeezed coherent states, quasi-free states and ground states of some free Hamiltonians, are the basis for various continuous variable quantum information processing \cite{Palma2017,Holevo2020,Srikara2020}. Since Gaussian states can be completely characterized by their first and second moments and the Gaussian character can be preserved under typical quantum operations, they are relatively easy to be studied analytically and are often used as theoretical testing ground \cite{Bondurant1984,Erkmen2008,Yokoyama2013,Chabaud2021}. Besides,  protocols based on Gaussian states and Gaussian measurements (e.g., homodyne detection) are relatively experimentally friendly as compared to operations of general quantum states \cite{Pinel2013,Banchi2015,McDonald2020,Bao2021}. Thus Gaussian states are of great importance in both theoretical studies and experimental implementations.

The preparation of a desired Gaussian state is clearly a pivotal task in continuous variable quantum information processing \cite{Diehl2008,Verstraete2009,Krauter2011,Koga2012,Yamamoto2012,S.Ma2014,S.Ma2019,ffner2020,Giovanni2021}. There have been some results proposed on the basis of the dissipation-induced approach or the environment engineering approach \cite{Diehl2008,Verstraete2009,Krauter2011,Koga2012,Yamamoto2012,S.Ma2014}. The basic idea behind those results is that one can utilize the dissipative environment by engineering a proper Hamiltonian and appropriately designed dissipative channels such that the corresponding stable state is a desired useful Gaussian state. In \cite{Koga2012}, necessary and sufficient conditions for a Gaussian master equation to have a unique steady pure state were  provided and then based on those conditions a systematic procedure to prepare a desired state via dissipation was proposed. Based on the quantum stochastic differential equation, \cite{Yamamoto2012} further investigated the pure Gaussian state generation and clarified a physical meaning that the nullifier dynamics of any Gaussian system generating a unique steady pure state is passive. An alternative method of preparing pure Gaussian state was presented in \cite{S.Ma2014}, where it was shown that a desired pure Gaussian state can be prepared by a cascade of one-dimensional open quantum harmonic oscillators, without any direct interaction Hamiltonians between these oscillators.

As mentioned, Gaussian states can be completely characterized by their first and second moments. In practice, most of the useful properties of Gaussian states such as the purity \cite{Paris2003}, entanglement \cite{Marian2008} and squeezing \cite{Petersen2005} are only related with the second moment. In previous results,  the covariance matrix which depicts the second moment of the Gaussian states is described in terms of the quantum master equation (QME) in essence \cite{Koga2012,Yamamoto2012,S.Ma2014}. In this paper, we provide an alternative way in preparing quantum Gaussian states by utilizing continuous measurement. To account for the randomness of  quantum measurement, the evolution of the open quantum system is described in terms of the quantum stochastic master equation (QSME). The basic idea is that the covariance matrix determined by the QSME is smaller (in the sense of matrix partial order) than that determined by the QME if all the parameterizations are the same. Thus,  by utilizing continuous measurement, the useful properties of  Gaussian states can be improved. We first present necessary and sufficient conditions for the system to have a unique stabilizing steady pure Gaussian state. The conditions are much weaker than those presented in the dissipation-induced approach. Then general parametric conditions of generating an arbitrarily desired pure Gaussian state are provided. They provide  more degrees of freedom to choose the system Hamiltonian and coupling channels as compared with previous results. For completeness and practical applications, we also consider the case where the detection is imperfect.

The paper is organized as follows. In Section II, we first give some preliminaries on  Gaussian states and the QSME, and then set up the linear quadratic Gaussian system model to be utilized. The main results focusing on pure Gaussian states are presented in Section III. Section IV discusses the case where the detection efficiency is less than 1. Section V concludes the paper.

\section{Gaussian states, QSME \& linear quadratic Gaussian system}

This section  briefly introduces the Wigner representation of Gaussian states, the framework of continuous measurement and the setting of linear quadratic Gaussian system. After an appropriate setting is given, the preparation problem of Gaussian states is restated as a stabilizing steady state problem of a Riccati  equation.

\subsection{Wigner representation of Gaussian states}\label{Section-G}

We start from  reviewing the Wigner phase-space representation \cite{Paris2003,Wiseman2010}, which is one of the most commonly used representations for quantum Gaussian states.

Denote $\hat{a}_k$ and $\hat{a}^\dag_k$, $k=1,...,m$, as the canonical bosonic annihilation and creation operators of the $k$-th mode, respectively. The corresponding canonical quadrature operators $\hat{q}_k$ and $\hat{p}_k$ are defined via $$\hat{q}_k=\frac{\hat{a}_k+\hat{a}_k^\dagger}{2}~~ \text{and}~~ \hat{p}_k=i\frac{\hat{a}_k^\dagger-\hat{a}_k}{2},$$ which satisfy the canonical commutation relation $$[\hat{q}_k,\hat{p}_j]=i\delta_{kj},$$ where $i=\sqrt{-1}$ and $\delta_{kj}$ is the Kronecker delta function.  Denote the vector of  quadrature operators for the $m$ mode system as  $$\hat{\textbf{X}}=(\hat{q}_1,\cdots,\hat{q}_m,\hat{p}_1,\cdots,\hat{p}_m)^\top,$$ where $\top$ is the matrix transpose. The operator vector $\hat{\textbf{X}}$ satisfies the canonical commutation relations \cite{Koga2012,Yamamoto2012,S.Ma2014}
\begin{equation}\nonumber
\left[\hat{\textbf{X}},\hat{\textbf{X}}^\top\right]=\hat{\textbf{X}}\hat{\textbf{X}}^{\top}-(\hat{\textbf{X}}\hat{\textbf{X}}^\top)^\top=iJ,
\end{equation}
where $J=\begin{pmatrix}
     0 & I_m \\
     -I_m & 0
\end{pmatrix}$, and $I_m$ is the $m\times m$ identity matrix.
%displacement operator of the $k$-th mode is described as $\hat{D}_k(\gamma)=\exp(\gamma \hat{a}^\dagger_k-\bar{\gamma}\hat{a}_k)$, where $\bar{\gamma}$ represents the conjugate of the complex number $\gamma$.
%
%The joint displacement operator on $d$ mode is
%\begin{equation}\nonumber
%\hat{D}(\boldsymbol{\gamma})=\hat{D}_1(\gamma_1)\otimes\cdot\cdot\cdot\otimes\hat{D}_n(\gamma_d),
%\end{equation}
%where $\boldsymbol{\gamma}=(\gamma_1,...,\gamma_d)$. And the system may be described by the quadrature field operators $\{\hat{q}_k,\hat{p}_k\}^d_{k=1}$ with $\hat{q}_k=\frac{\hat{a}_k+\hat{a}_k^\dagger}{2}$ and $\hat{p}_k=i\frac{\hat{a}_k^\dagger-\hat{a}_k}{2}$, which can be formally arranged in the vector $$\hat{\xi}:=(\hat{q}_1,\hat{p}_1,...,\hat{q}_d,\hat{p}_d)^{T}.$$ The canonical commutation relations can then be written
%$$
%[\hat{\xi}_i,\hat{\xi}_j]=i\Omega_{ij},
%$$
%where $\Omega$ is the symplectic matrix defined by
%$$
%\Omega=\mathop{\bigoplus}\limits_{k=1}^{d}\begin{pmatrix}
%     0 & 1 \\
%     -1 & 0
%\end{pmatrix}.
%$$
%By using the vector of quadrature operators and the symplectic matrix above,

The joint displacement operator can be described in terms of $\hat{\textbf{X}}$ and $J$ as
$$\hat{D}(\boldsymbol{\alpha})=\textrm{exp}(i~\hat{\textbf{X}}^{\top}J\boldsymbol{\alpha}),$$
where $\boldsymbol{\alpha}\in\mathbb{R}^{2m}$ \cite{Paris2003,Wiseman2010}. On the basis of $\hat{D}(\boldsymbol{\alpha})$, for an arbitrary quantum state operator $\hat{\rho}$, its  Wigner characteristic function is defined as
\begin{equation}\nonumber
\chi_{\hat{\rho}}(\boldsymbol{\alpha})=\textrm{Tr}[\hat{\rho}\hat{D}(\boldsymbol{\alpha})].
\end{equation}
Now the Wigner function $W_{\hat{\rho}}(X)$ of the state $\hat{\rho}$ can  be described as
\begin{equation}\nonumber
W_{\hat{\rho}}(X)=\int_{\mathbb{R}^{2m}}\frac{d\boldsymbol{\alpha}}
{(2\pi)^{2m}}\textrm{exp}(-iX^{\top}J\boldsymbol{\alpha})\chi_{\hat{\rho}}(\boldsymbol{\alpha}),
\end{equation}
where  $X=(q_1,...,q_m, p_1, \cdots,p_m)^{\top}$ with $q_i$ and $p_j$ being real numbers \cite{Paris2003,Wiseman2010}.

A state is called  Gaussian if its Wigner function is in the form of
\begin{equation}\label{wigner}
W(X)=\frac{\textrm{exp}[-\frac{1}{2}(X-\bar{X})^{\top}V^{-1}(X-\bar{X})]}{(2\pi)^{m}\sqrt{\textrm{det}[V]}},
\end{equation}
where $\bar{X}=\langle\hat{\textbf{X}}\rangle=\textrm{Tr}(\hat{\textbf{X}}\hat{\rho})$ is the mean value vector whose element $\bar{X}_i=\langle\hat{\textbf{X}}_i\rangle=\textrm{Tr}(\hat{\textbf{X}}_i\hat{\rho})$,   $V$ is the covariance matrix whose element  $V_{ij}=\frac{1}{2}\langle \Delta\hat{\textbf{X}}_i\Delta\hat{\textbf{X}}_j+\Delta\hat{\textbf{X}}_j \Delta\hat{\textbf{X}}_i\rangle$  with $\Delta\hat{\textbf{X}}_i=\hat{\textbf{X}}_i-\langle\hat{\textbf{X}}_i\rangle$, and $\textrm{det}[V]$ denotes the determinant of matrix $V$  \cite{Edwards2005,Wiseman2010}. In addition to the positive semidefinite condition $V\geq 0$, $V$ should satisfy the Heisenberg uncertainty principle \cite{Koga2012,Yamamoto2012,S.Ma2014}
\begin{equation*}\label{heisenberg}
V\geq \pm\frac{i}{2}J.
\end{equation*}
It is straightforward to see that a Gaussian state is determined  by the mean vector $\bar{X}$ and the covariance matrix $V$.

In this paper, we only focus on the covariance matrix $V$. This is because, on the one hand, that one can always adjust the mean of a Gaussian state to any target mean by a suitable Weyl operator \cite{Parthasarathy2010}. On the other hand it is noted that most of the useful properties of Gaussian states such as the purity, entanglement and squeezing are only related with the covariance matrix in practice.  To be specific,  the purity of a $m$ mode Gaussian state depends only on the covariance matrix as $\textrm{Tr}[\hat{\rho}^2]=\frac{1}{2^{m}\sqrt{\textrm{det}[V]}}$ \cite{Paris2003}. Moreover, for a two-mode pure Gaussian state, its $4\times4$ covariance matrix contains all the necessary information to determine its entanglement properties for both entanglement criteria and entanglement measures \cite{Rendell2005,Marian2008}. In addition, squeezed coherent states have been extensively studied in quantum information, such as quantum teleportation and ghost imaging. The squeezing properties are also determined by the covariance matrix \cite{Petersen2005}.

\subsection{Quantum stochastic master equation}

In previous results, the commonly used model in preparing a target Gaussian state is to engineer a dissipative system described by the quantum master equation \cite{Edwards2005,Wiseman2010,Koga2012,Yamamoto2012}
\begin{equation}\label{QME}
\frac{d\hat{\rho}_t}{dt}=-i[\hat{H},\hat{\rho}_t]+\sum_{i=1}^{m}(\hat{L}_i \hat{\rho}_t \hat{L}_i^\dagger-\frac{1}{2}\hat{L}_i^\dagger \hat{L}_i \hat{\rho}_t-\frac{1}{2}\hat{\rho}_t\hat{L}_i^\dagger \hat{L}_i ),
\end{equation}
where $\hat{\rho}_t$ is the density operator, $\hat{H}$ is the effective Hamiltonian of the system and $\hat{L}_i,~ i=1,\ldots,m$, are the dissipative channels which describe the interaction between the system and the  environment. The dissipative QME (\ref{QME}) can be viewed as an average (unconditional) evolution of the quantum state under the assumption that all the measurement results are discarded. Intuitively if one can utilize the measurement information, the corresponding conditional covariance matrix may be reduced, and thus the useful properties of the Gaussian states can be improved.
In this paper, we present how to prepare a desired quantum Gaussian state by utilizing continuous measurement. To account for the randomness of  quantum measurement, the evolution of the open quantum system is described in terms of the quantum stochastic master equation.

To proceed, we specify the following notation for clarity.  For a matrix $Q=(q_{ij})$,  let $Q^\dagger$, $Q^{\top}$ and $Q^*$ represent the conjugate transpose, transpose and  conjugate for all the elements of $Q$, respectively. For example, $Q^\dagger=(q^*_{ji}),~Q^{\top}=(q_{ji}),$ and $Q^*=(q^*_{ij})=(Q^\dagger)^{\top}$.

The dynamics of the open system coupled with $m$ independent measurement channels $\{\hat{L}_i\}^m_{i=1}$ can be described by a family of unitary operators $\{\hat{U}_t\}_{t\in\mathbb{R_+}}$, which  satisfy
\begin{equation}\label{U}
\begin{aligned}
d&\hat{U}_t=-\hat{K}\hat{U}_t dt+\sum^m_{i=1}(\hat{L}_id\hat{A}^\dagger_{i,t}-\hat{L}_i^\dagger d\hat{A}_{i,t})\hat{U}_t,\\
&\hat{U}_0=I,
\end{aligned}
\end{equation}
where $\hat{K}=i\hat{H}+\frac{1}{2}\sum^m_{i=1}\hat{L}^\dagger_i \hat{L}_i$ \cite{Edwards2005}. Here we have assumed $\hbar=1$. The differential of the annihilation operators of the bath $d\hat{A}_{i,t}$ obeys the quantum It$\hat{\textrm{o}}$ rules:
$$d\hat{A}_{i,t}d\hat{A}_{j,t}^{\dagger}=\delta_{i,j}dt,~d\hat{A}_{i,t}d\hat{A}_{j,t}=0,$$
$$d\hat{A}^{\dagger}_{i,t}d\hat{A}^{\dagger}_{j,t}=0,~d\hat{A}^{\dagger}_{i,t}d\hat{A}_{j,t}=0.$$

For any system operator $\hat{X}$, its time evolution in the Heisenberg picture is represented by  $$j_t(\hat{X})=\hat{U}_t^\dagger (\hat{X}\otimes I)\hat{U}_t.$$  From  (\ref{U}) and the quantum   It$\hat{\textrm{o}}$ rules, we have
\begin{equation}\label{jtX}
\begin{aligned}
dj_t(\hat{X})=j_t(\mathcal{L}[\hat{X}])dt+&\sum^m_{i=1}\Big{(}j_t([\hat{L}^{\dagger}_{i},\hat{X}])dA_{i,t}
\\&+j_t([\hat{X},\hat{L_i}])dA^{\dagger}_{i,t}\Big{)},
\end{aligned}
\end{equation}
where $$\mathcal{{L}}[\hat{X}]=i[\hat{H},\hat{X}]+\sum_{i=1}^{m}(\hat{L}_i^\dagger \hat{X} \hat{L}_i-\frac{1}{2}\hat{L}_i^\dagger \hat{L}_i \hat{X}-\frac{1}{2}\hat{X}\hat{L}_i^\dagger \hat{L}_i ).$$

Consider $m$ family of measurement operators $\{Y_{i,t}\}_{t\in\mathbb{R_+}}$,  where $$\hat{Y}_{i,t}=\hat{U}^\dagger_t(\hat{A}_{i,t}+\hat{A}^\dagger_{i,t})\hat{U}_t$$ for $i=1,\cdots,m$. From   (\ref{U}) and the quantum   It$\hat{\textrm{o}}$ rules, we have
\begin{equation}\label{Yt}
d\hat{Y}_{i,t}=j_t(\hat{L}_i+\hat{L}_i^\dagger)dt+d\hat{A}_{i,t}+d\hat{A}^\dagger_{i,t}.
\end{equation}
Once $\hat{Y}_{i,t}=\hat{U}^\dagger_t(\hat{A}_{i,t}+\hat{A}^\dagger_{i,t})\hat{U}_t$, $i=1, \cdots, m$, are observed, one can associate a conditional expectation to each observable $\hat{X}$ of the system. The  conditional expectation gives the least squares estimator of $j_t(\hat{X})$ as $\mathcal{E}[j_t(\hat{X})| \hat{Y}^t_0]$, where $\hat{Y}^t_0$ is the algebra generated by $\hat{Y}_{i,s\leq t}$, $i=1,\cdots,m$. Since $\hat{Y}^t_0$ is nondemolition, the observation process is in essence equivalent to $m$ classical stochastic processes $$\{y_{i,s\leq t}, i=1,\cdots, m\}.$$ Thus $\mathcal{E}[j_t(\hat{X})| \hat{Y}^t_0]$ actually represents the expectation of $\hat{X}$ conditioned on the classical stochastic observations $\{y_{i,s\leq t}$, $i=1,\cdots, m\}$.  This conditional expectation can be conveniently written in the Schr\"{o}dinger picture as $$\mathcal{E}[j_t(\hat{X})| \hat{Y}^t_0]=\text{Tr}(\hat{\rho}^c_t \hat{X})$$ with $\hat{\rho}^c_t $ denoting the conditional system state at time $t$. The stochastic master equation of $\rho^c_t$ is described as \cite{Handel2005,Edwards2005,Wiseman2010}
\begin{equation}\label{conditionstate}
\begin{aligned}
d\hat{\rho}_t^c=\mathcal{L}^\dagger[\hat{\rho}^c_t]dt+&\sum_{i=1}^m\Big{(}\hat{L}_i\hat{\rho}^c_t +\hat{\rho}^c_t \hat{L}^\dagger_i\\&-\text{Tr}[\hat{\rho}^c_t (\hat{L}_i+\hat{L}_i^\dagger)]\hat{\rho}^c_t \Big{)}dw_{i,t},
\end{aligned}
\end{equation}
where the innovations $$dw_{i,t}=d{y}_{i,t}-\langle\hat{\rho}^c_t ,\hat{L}_i+\hat{L}_i^\dagger\rangle dt$$ are Gaussian, and
$$\mathcal{{L}}^\dagger[\hat{\rho}_t^c]=-i[\hat{H},\hat{\rho}_t^c]+\sum_{i=1}^{m}(\hat{L}_i
 \hat{\rho}_t^c \hat{L}_i^\dagger-\frac{1}{2}\hat{L}_i^\dagger \hat{L}_i \hat{\rho}_t^c-\frac{1}{2}\hat{\rho}_t^c\hat{L}_i^\dagger \hat{L}_i ).$$
It is clear that by taking the expectation of \eqref{conditionstate}, we can obtain the quantum master equation \eqref{QME}.

\subsection{Quantum linear quadratic Gaussian system}

We are interested  in systems of $m$ degrees of freedom with the phase space operator vector  $$\hat{\textbf{X}}=(\hat{q}_1,\cdots,\hat{q}_m,\hat{p}_1,\cdots,\hat{p}_m)^\top$$ as defined in subsection 2.1. Recall that
the operator vector $\hat{\textbf{X}}$ satisfies the canonical commutation relation
\begin{equation}\nonumber
\begin{aligned}
\left[\hat{\textbf{X}},\hat{\textbf{X}}^\top\right]&=\hat{\textbf{X}}\hat{\textbf{X}}^{\top}-(\hat{\textbf{X}}\hat{\textbf{X}}^\top)^\top=iJ.\\
\end{aligned}
\end{equation}
The system is coupled to $m$ measurement channels via the operator vector $\hat{\textbf{L}}=\Lambda \hat{\textbf{X}}$, where $\Lambda\in \mathbb{C}^{m \times 2m}$. The  Hamiltonian which is quadratic in $\hat{\textbf{X}}$ is represented as
\begin{equation}\nonumber
\hat{H}(u_t)=\frac{1}{2}\hat{\textbf{X}}^\top G ~{\hat{\textbf{X}}} +\hat{\textbf{X}}^\top K u_t+u_t^\top K^\dagger \hat{\textbf{X}},
\end{equation}
where $G=G^\top \in \mathbb{R}^{2m\times2m}$, $K\in \mathbb{C}^{2m\times m}$ and $u_t\in \mathbb{R}^m$ is the control. For such a system,
from  (\ref{U})-(\ref{Yt}), we have the quantum linear equations in the Heisenberg picture as
\begin{equation*}\label{quantum}
\begin{aligned}
&d\hat{\textbf{X}}_t=(A\hat{\textbf{X}}_t+Bu_t)dt+d\hat{\textbf{V}}_t,\\
&d\hat{\textbf{Y}}_t=C\hat{\textbf{X}}_tdt+d\hat{\textbf{W}}_t,
\end{aligned}
\end{equation*}
where
\begin{equation*}
 \begin{aligned}
A&=J\Big{(}G+\frac{1}{2i}(\Lambda^\dagger \Lambda-\Lambda^\top \Lambda^*)\Big{)}\in \mathbb{R}^{2m\times2m},\\
B&=J(K+K^*)\in \mathbb{R}^{2m\times m},\\
C&=\Lambda+\Lambda^*\in \mathbb{R}^{m\times 2m},
\end{aligned}
\end{equation*}
 and the quantum noise increments are given by
\begin{equation*}\label{quantum-noise}
\begin{aligned}
&d\hat{\textbf{V}}_t=iJ(\Lambda^\top d\hat{\textbf{A}}_t^\dagger-\Lambda^\dagger d\hat{\textbf{A}}_t),\\
&d\hat{\textbf{W}}_t=d\hat{\textbf{A}}_t+d\hat{\textbf{A}}_t^\dagger.
\end{aligned}
\end{equation*}

For the quantum linear quadratic system, if the initial state $\hat{\rho}$ of the system is Gaussian, then under the nondemolition measurement of the output operators $\hat{\textbf{Y}}_t$, the conditional dynamics (\ref{conditionstate}) is also Gaussian. Thus for the linear quadratic Gaussian system, its dynamics can be sufficiently described by the time flow of the conditional mean vector $\bar{X}_t=\textrm{Tr}(\hat{\textbf{X}}\hat{\rho}^c_t)$  and the covariance matrix $V_t$, whose element  $$V_t{_{,ij}}=\frac{1}{2}\langle \Delta_c\hat{\textbf{X}}_i\Delta_c\hat{\textbf{X}}_j+\Delta_c\hat{\textbf{X}}_j \Delta_c\hat{\textbf{X}}_i\rangle_c$$ with $\Delta_c\hat{\textbf{X}}_i=\hat{\textbf{X}}_i-\langle\hat{\textbf{X}}_i\rangle_c$ and $\langle \hat{\textbf{X}}_i \rangle_c=\text{Tr}(\hat{\textbf{X}}_i\rho^c_t)$. Using  (\ref{conditionstate}), we have the conditional expectation $\bar{X}_t$ of $\hat{\textbf{X}}$ under the measurement of $\hat{\textbf{Y}}_t$ as
\begin{equation}\label{conditionexpectation}
\begin{aligned}
d\bar{X}_t&=(A\bar{X}_t+Bu_t)dt+(V_t C^\top+M)d\tilde{y}_t,\\
\bar{X}_0&=\bar{X},
\end{aligned}
\end{equation}
where $$M=\frac{i}{2}J(\Lambda^\top-\Lambda^\dagger)\in \mathbb{R}^{2m\times m},$$ and $$d\tilde{y}_t=dy_t-C\bar{X}_tdt$$ is the Gaussian innovation which describes the information gain from the the measurement. The conditional covariance matrix satisfies the matrix Riccati equation
\begin{equation}\label{conditioncovariance}
\begin{aligned}
\frac{d}{dt}V_t&=AV_t+V_tA^\top+N-(V_t C^\top+M)(V_t C^\top+M)^\top,\\
V_0&=V,
\end{aligned}
\end{equation}
where $$N=\frac{1}{2}J(\Lambda^\dagger \Lambda+\Lambda^\top \Lambda^*)J^\top\in \mathbb{R}^{2m\times 2m}.$$
It is worth noting that the equation for $V_t$ is deterministic and does not depend on the stochastic measurement results.

Taking expectation values of \eqref{conditionexpectation} and \eqref{conditioncovariance}, we have the moment equations for the unconditional approach as
\begin{equation}\label{uncondition}
\begin{aligned}
&d\bar{X}_t=(A\bar{X}_t+Bu_t)dt,\\
&\frac{d}{dt}V_t=AV_t+V_tA^\top+N,
\end{aligned}
\end{equation}
which correspond to the quantum master equation (\ref{QME}). To prepare a desired Gaussian state, most of the previous results are on the basis of the moment equations (\ref{uncondition}). Note that the final term in (\ref{conditioncovariance}) causes a reduction in the conditional covariance matrix as compared with that in  (\ref{uncondition}). Recall that most of the useful properties of the Gaussian states are determined by the covariance matrix. Thus intuitively by utilizing the conditional covariance matrix equation (\ref{conditioncovariance}), one may provide weaker conditions in preparing a desired Gaussian state.

%As it's written, sometimes these conditions are a little bit cumbersome. The quantum stochastic master equation method gives a more concise condition. Next we will explore the preparation of Gaussian pure states by the quantum random master equation method.

Let us first look at the equation of the conditional expectation (\ref{conditionexpectation}). We can choose $B$ such that  its column space  is the same as that of $VC^\top+M$, where $V$ is the steady solution of  (\ref{conditioncovariance}). Then we  choose $F$ such that it satisfies $$BF=-V C^\top-M.$$ Thus, for the Markovian feedback (or direct feedback) $u_tdt=F dy_t$ \cite{Wiseman2010}, the equation \eqref{conditionexpectation} of the first moment becomes deterministic in the limit $t\rightarrow\infty$ as
\begin{equation}\label{controlequation}
\frac{d}{dt}\bar{X}_t=(A-MC-VC^\top C)\bar{X}_t.
\end{equation}
Therefore, if $A-MC -VC^\top C$ is stable, then the solution of \eqref{controlequation} will approach to 0 approximately. In the following we will give  necessary and sufficient conditions for  the conditional covariance matrix equation \eqref{conditioncovariance} to have a  steady solution and meanwhile $A-MC-VC^\top C$ is stable. From now on we focus on the conditional covariance matrix equation \eqref{conditioncovariance}.

\section{Main results}
In subsection 3.1, we first present necessary and sufficient condition for the Riccati equation (8) to have a unique stabilizing steady solution, which corresponds to a conditional covariance matrix of some Gaussian state. In subsection 3.2, we further prove that the corresponding Gassian state is pure. Then in subsection 3.3, we provide algebraic characterization of how to prepare a pure Gaussian state with the desired covariance matrix.

\subsection{Stabilizing steady solution of the Riccati equation}

Above all, we present the following lemma (Theorem 3.4 in \cite{zhou2002}) concerning the properties of detectability, which is to be utilized in the main results.
\begin{lemma}
The following are equivalent:\\
(i) $[C,~A]$ is detectable;\\
(ii) The matrix $\begin{bmatrix}
     A-\lambda I  \\
     C
\end{bmatrix}$ has full column rank for all $\textrm{Re}(\lambda) \geq0$;\\
(iii) For all $\lambda$ and $x\neq0$ such that $Ax=\lambda x$ and $\textrm{Re}(\lambda)\geq0$, $Cx\neq0$;\\
(iv) There exists a matrix $F$ such that $A + FC$ is stable, i.e., all the eigenvalues of $A+FC$ have negative real parts;\\
(v) $(A^\top,~C^\top)$ is stabilizable.
\end{lemma}
The definition of stabilizing steady solution is defined as the following.

\begin{definition}
A matrix $V$ is called a \textit{stabilizing steady }solution of \eqref{conditioncovariance}, if

(i) it satisfies the algebraic  Riccati equation
\begin{equation*}\label{AR}
AV+VA^\top+N-(V C^\top+M)(V C^\top+M)^\top=0,
\end{equation*}
or, equivalently,
\begin{equation}\label{riccati}
(A-MC)V+V(A-MC)^\top-VC^\top CV+\frac{1}{4}JC^\top CJ^\top =0,
\end{equation}
where $$A-MC=J\Big{(}G+\frac{1}{2i}(\Lambda^\top \Lambda-\Lambda^\dagger \Lambda^*)\Big{)}\in\mathbb{R}^{2m\times 2m}.$$

(ii) $A-MC-VC^\top C$ is stable, i.e., all the eigenvalues of $A-MC-VC^\top C$ sit in the left half plane.
\end{definition}
Now we provide necessary and sufficient conditions for the Riccati equation (\ref{conditioncovariance}) to have a unique stabilizing steady solution, which corresponds to a  quantum Gaussian state.

\begin{thm}\label{thm1}
The Riccati equation \eqref{conditioncovariance} has a unique stabilizing steady solution $V$ which is  positive semidefinite if and only if  $[C,~A]$ is detectable.
\end{thm}

%
%To prove Theorem \ref{thm1}, we first introduce the following two lemmas.
%
%Suppose $F\in\mathbb{R}^{n\times n},~O\in\mathbb{R}^{n\times m},$ and $Z\in\mathbb{R}^{k\times n}$. Consider the following Riccati equation concerning $X$
%\begin{equation}\label{r2}
%F^\top X+XF-XOO^\top X+Z^\top Z=0.
%\end{equation}
%A solution $X$ is called\textit{ stabilizing} if all the eigenvalues of $F-OO^\top X$ sit in the left half plane.
%
%According to Theorem 13.7 in \cite{zhou2002}, we have Lemma 1.
%\begin{lemma}\label{lemma1}
%The Riccati equation (\ref{r2}) has a unique  stabilizing solution  which is positive semidefinite if
%
%(i) $[O^\top,~F^\top]$ is detectable;
%
%(ii) $[Z,~F]$ has no unobservable modes on the imaginary axis.
%\end{lemma}
%\begin{remark}
%In general, the two conditions in Lemma \ref{lemma1} are independent. However,  owing to the specific structure in \eqref{riccati}, the above two conditions in proving Theorem 1 can be simplified.
%\end{remark}
%\begin{lemma}\label{lemma2}
%$[CJ^\top,~(A-MC)^\top]$ has no unobservable modes on the imaginary axis if $[C,~A-MC]$ is detectable.
%\end{lemma}
%
%The proof of Lemma 2 is presented in Appendix B. Now we give the proof of Theorem \ref{thm1}.

{\bf  Proof:} \textit{Sufficiency:} Firstly, according to Theorem 13.7 in \cite{zhou2002}, if
\begin{enumerate}
\item $[C,~A-MC]$ is detectable, and
\item $[CJ^\top,~(A-MC)^\top]$ has no unobservable modes on the imaginary axis,
\end{enumerate}
then the Riccati equation (\ref{riccati}) has a unique stabilizing solution $V$, i.e., $A-MC-VC^\top C$ is stable. Moreover,  $V$ is positive semidefinite.

Secondly, let us prove that  $[CJ^\top,~(A-MC)^\top]$ has no unobservable modes on the imaginary axis if  $[C,~A-MC]$ is detectable.
Suppose, on the contrary, that $[CJ^\top,~(A-MC)^\top]$ has an unobservable mode on the imaginary axis. From Lemma 1, there exists $w\in \mathbb{R}$ and a corresponding $x\neq 0$, such that
\begin{equation}\nonumber
\begin{aligned}
(A-MC)^\top x&=iwx,\\
CJ^\top x&=0.
\end{aligned}
\end{equation}
Thus, we have
\begin{equation}\nonumber
\begin{aligned}
(A-MC)J^\top x&=J (G+\frac{1}{2i}(\Lambda^\top  \Lambda-\Lambda^\dagger \Lambda^*))J^\top x\\
&=J(A-MC)^\top x\\
&=Jiw x\\
&=-iwJ^\top x
\end{aligned}
\end{equation}
Combining this conclusion with $CJ^\top x=0$ contradicts with the condition that  $[C,~A-MC]$ is detectable. Thus, $[CJ^\top,~(A-MC)^\top]$ has no unobservable modes on the imaginary axis if $[C,A-MC]$ is detectable.  Moreover, it is straightforward to verify that $[C,~A-MC]$ being detectable is equivalent to $[C,~A]$ being detectable.

\textit{Necessity:} Since the unique stabilizing solution of the Riccati equation \eqref{riccati} can be represented as \cite{zhou2002}
\begin{equation}\nonumber
\begin{aligned}
V=&\int_0^{\infty}\textrm{e}^{\big{(}(A-MC)^\top-C^\top CV\big{)}^\top t}~\big{(}VC^\top CV\\
&+\frac{1}{4}JC^\top CJ^\top\big{)}~\textrm{e}^{\big{(}(A-MC)^\top -C^\top CV\big{)}t}dt,
\end{aligned}
\end{equation}
the matrix $(A-MC)^\top -C^\top CV$ should be stable. This implies that  $[(A-MC)^\top,~C^\top]$ is stabilizable,  and  $[C,~A]$ is detectable accordingly.~~~$\blacksquare$

\begin{remark}
A similar result was also given in \cite{Wiseman2010}, while our proof is different from that there. In existing results, to ensure the system described by  (\ref{uncondition}) to have a unique steady state, the system matrix $A$ is usually supposed to be stable. However, under the continuous measurement and quantum filtering framework, Theorem \ref{thm1} provides a much weaker condition for the Riccati equation concerning the conditional covariance matrix to have a unique stabilizing steady solution.
\end{remark}

It can be seen that if $[C,~A]$ is detectable, $A-MC -VC^\top C$  is stable, and the solution of \eqref{controlequation} will approximately approach to 0.

From Theorem 1, if $[C,~A]$ is detectable, then there is a unique stabilizing solution of the Riccati equation \eqref{riccati} satisfying $V\geq0$. But as a covariance matrix depicting quantum Gaussian state, $V$ must satisfy the Heisenberg uncertainty principle, $$V\geq\pm\frac{i}{2}J.$$ In fact,  $V\geq\pm\frac{i}{2}J$ essentially implies $V> 0$ as proved in \cite{Pirandola2009}.
%
%\begin{remark}
%V is a real symmetric matrix and $V>0$ if $V\geq\pm\frac{i}{2}J$.
%\end{remark}

A necessary and sufficient condition for the solution $V$ of the Riccati equation \eqref{riccati} to satisfy the Heisenberg uncertainty principle is given in Theorem 2.

\begin{thm}
The Riccati equation \eqref{conditioncovariance} has a unique  stabilizing steady solution $V$ satisfying $V\geq\pm\frac{i}{2}J$ if and only if  $[C,~A]$ is detectable.
\end{thm}

Since the proof of Theorem 2 involves a Riccati equation in the complex field. We introduce two lemmas for Riccati equations in the complex field.
\begin{lemma}\label{lemma3}
Suppose $F\in\mathbb{C}^{n\times n},~O\in\mathbb{C}^{n\times m},$ and $Z\in\mathbb{C}^{k\times n}$.  If
$[O^\dagger,~F^\dagger]$ is detectable and $[Z,F]$ has no unobservable modes on the imaginary axis, then the Riccati equation for $X$
\begin{equation}\nonumber
F^\dagger X+XF-XOO^\dagger X+Z^\dagger Z=0
\end{equation}
has a unique stabilizing solution which is positive semidefinite.
\end{lemma}

The proof of Lemma \ref{lemma3} is presented in Appendix B by analyzing the Riccati equation in the complex domain (see Appendix A).

\begin{lemma}\label{lemma4}
$[CJ^\top,~(A-MC)^\top]$ has no unobservable modes on the imaginary axis if and only if $A-MC-\frac{i}{2}JC^TC$ has no eigenvalues on the imaginary axis.
\end{lemma}

{\bf  Proof:}
\textit{Sufficiency:} Suppose, on the contrary, that  $[CJ^\top ,~(A-MC)^\top ]$ has an unobservable mode on the imaginary axis. Then there exists $\lambda\in \mathbb{R}$ and a corresponding vector $x\neq 0$, such that
\begin{equation}\nonumber
\begin{aligned}
(A-MC)^\top x&=i\lambda x,\\
CJ^\top x&=0.
\end{aligned}
\end{equation}
Thus, we have
\begin{equation}\nonumber
\begin{aligned}
&(A-MC-\frac{i}{2}JC^\top C)^\top x\\
=&(A-MC)^\top x-\frac{i}{2}C^\top CJ^\top x\\
=&i\lambda x,
\end{aligned}
\end{equation}
which contradicts with the condition that $A-MC-\frac{i}{2}JC^\top C$ has no eigenvalues
on the imaginary axis.

\textit{Necessity:} Suppose $A-MC-\frac{i}{2}JC^\top C$ has an eigenvalue
on the imaginary axis. Then there exists $\lambda\in \mathbb{R}$ and a corresponding vector $x\neq 0$, such that
\begin{equation}\label{eq-lemma4}
(A-MC-\frac{i}{2}JC^\top C)x=i\lambda x.
\end{equation}
Multiplying the above equation from left by $x^\dagger J^\top $ yields
\begin{equation}\nonumber
\begin{aligned}
&x^\dagger J^\top(A-MC-\frac{i}{2}JC^\top C)x\\
=&x^\dagger\big{(}G+\frac{1}{2i}(\Lambda^\top  \Lambda-\Lambda^\dagger \Lambda^*)\big{)}x-\frac{i}{2}x^\dagger C^\top Cx,\\
=&i\lambda x^\dagger J^\top x.
\end{aligned}
\end{equation}
Since $(x^\dagger Jx)^\dagger=-x^\dagger Jx$ is a pure imaginary number and $G+\frac{1}{2i}(\Lambda^\top  \Lambda-\Lambda^\dagger \Lambda^*)$ as well as $C^\top C$ are real symmetric matrices, we have $Cx=0$, and $(A-MC)x=i\lambda x$ from (\ref{eq-lemma4}), accordingly. Then, since $(A-MC)J^\top=J(A-MC)^\top$, we have
\begin{equation}\nonumber
\begin{aligned}
(A-MC)^\top Jx&=J^\top (A-MC)J^\top Jx\\
&=J^\top i\lambda x\\
&=-i\lambda Jx,\\
CJ^\top Jx&=0,
\end{aligned}
\end{equation}
which contradicts with the condition that $[CJ^\top ,~(A-MC)^\top ]$ has no unobservable modes on the imaginary axis.

Thus, $[CJ^T,~(A-MC)^T]$ has no unobservable modes on the imaginary axis if and only if $A-MC-\frac{i}{2}JC^TC$ has no eigenvalues on the imaginary axis.~~~$\blacksquare$

Now we give the proof of Theorem 2.

{\bf  Proof:} \textit{Sufficiency:} From Theorem 1, if  $[C,~A]$ is detectable, then the Riccati equation $(\ref{conditioncovariance})$ has a unique stabilizing steady solution $V$ which is positive semidefinite. Since $V$ is real and symmetric, and $J$ is antisymmetric, we just need to prove $V\geq\frac{i}{2}J$.

Suppose $Y=V-\frac{i}{2}J$. From  $(\ref{riccati})$ we have the Riccati equation for $Y$ as
\begin{equation}\label{riccatiY}
\begin{aligned}
(A-MC-&\frac{i}{2}JC^\top C)Y+Y(A-MC-\frac{i}{2}JC^\top C)^\dagger\\
&-YC^\top CY=0.
\end{aligned}
\end{equation}
From the proof of Theorem 1, if $[C,~A-MC]$ is detectable, then $[CJ^\top,~(A-MC)^\top]$ has no unobservable modes on the imaginary axis.
This conclusion combined with Lemma \ref{lemma4} implies that $A-MC-\frac{i}{2}JC^\top C$ has no eigenvalues on the imaginary axis. Thus, $[0,(A-MC-\frac{i}{2}JC^\top C)^\dagger]$ has no unobservable modes on the imaginary axis. Moreover, $[C,~A-MC-\frac{i}{2}JC^\top C]$ is detectable if  $[C,~A]$ is detectable. According to Lemma \ref{lemma3}, the Riccati equation \eqref{riccatiY} for $Y$ has a unique stabilizing solution which is positive semidefinite, i.e., $V\geq\frac{i}{2}J$, if $[C,~A]$ is detectable.

 \textit{Necessity:} %According to \cite{zhou2002}, the solution of \eqref{riccatiY} has the following form
%\begin{equation}\nonumber
%\begin{aligned}
%Y=\int_0^{\infty}&\textrm{e}^{((A-MC-\frac{i}{2}JC^\top C)^\top -C^\top CY)^\top t}YC^\top \\
%&CY\textrm{e}^{((A-MC-\frac{i}{2}JC^\top C)^\top -C^\top CY)t}.
%\end{aligned}
%\end{equation}
%Obviously $((A-MC-\frac{i}{2}JC^\top C)^\top -C^\top CY)$ is stable, that is, $[C,A-MC]$ is detectable.
It can be obtained from the necessity part of Theorem 1.~~~$\blacksquare$

\subsection{Stabilizing steady pure  Gaussian state}

From Theorem 2 we can see that if $[C,~A]$ is detectable, then the stabilizing steady state solution of \eqref{conditioncovariance} corresponds to a conditional covariance matrix of some Gaussian state. Since pure Gaussian states of continuous variable systems are the key ingredients of secure optical communications and Heisenberg limited interferometry, the characterization of pure Gaussian states has attracted much attention.

Note that the purity of a $m$ mode Gaussian state is  simply $$\textrm{Tr}[\hat{\rho}^2]=\frac{1}{2^{m}\sqrt{\textrm{det}[V]}}.$$
%For a pure Gaussian state,  the corresponding covariance matrix $V$ satisfies $JVJV=-\frac{I}{4}$. This is also hard to verify without having the specific form of the steady solution $V$.
In \cite{Koga2012}, by utilizing a dissipation-induced approach on the basis of the model (\ref{uncondition}), an explicit algebraic characterization of the purity of Gaussian state was given. For the model  (\ref{uncondition}), to  have a unique stabilizing steady pure state with covariance matrix $V_s$, besides $A$ being stable, $V_s$ must satisfy the following matrix equations:
 \begin{equation}\label{un1}
(V_s+\frac{i}{2}J)\Lambda^\top=0,
~JGV_s+V_sGJ^\top=0.
\end{equation}
However, these matrix equations are not easy to verify without knowing the solution $V_s$.

A natural question is whether the unique stabilizing steady solution $V$ of the Riccati equation \eqref{conditioncovariance} satisfying $V\geq\pm\frac{i}{2}J$ corresponds to some pure Gaussian state. The following Theorem 3 gives an affirmative answer.
\begin{thm}
If  $[C,~A]$ is detectable, then the Riccati equation \eqref{conditioncovariance} has a unique stabilizing steady solution $V$, which can be considered as a conditional covariance matrix of some pure Gaussian state.
\end{thm}

{\bf  Proof:} To prove the steady state being pure, we resort to the correspondence of  Gaussian states between the Schr$\ddot{\textrm{o}}$dinger picture and the Heisenberg picture.

Consider the following stochastic master equation of $\hat{\rho}_t$ starting from an initial pure Gaussian state $\hat{\rho}_0$,
\begin{equation}\label{conditionstate1}
d\hat{\rho}_t=\mathcal{L}^\dagger[\hat{\rho}_t]dt+\sum_{i=1}^m\Big{(}\hat{L}_i\hat{\rho}_t +\hat{\rho}_t \hat{L}^\dagger_i-\text{Tr}[\hat{\rho}_t (\hat{L}_i+\hat{L}_i^\dagger)]\hat{\rho}_t \Big{)}dw_{i,t},
\end{equation}
where the innovations $$dw_{i,t}=d{y}_{i,t}-\langle\hat{\rho}_t ,\hat{L}_i+\hat{L}_i^\dagger\rangle dt$$ are Gaussian, and
$$\mathcal{{L}}^\dagger[\hat{\rho}_t]=-i[\hat{H},\hat{\rho}_t]+\sum_{i=1}^{m}(\hat{L}_i
 \hat{\rho}_t \hat{L}_i^\dagger-\frac{1}{2}\hat{L}_i^\dagger \hat{L}_i \hat{\rho}_t-\frac{1}{2}\hat{\rho}_t\hat{L}_i^\dagger \hat{L}_i ).$$
It is straightforward to verify that
$$d~\text{Tr}(\hat{\rho}^2_t)=0.$$
This combined with the initial state $\hat{\rho}_0$ being pure implies that the state $\hat{\rho}_t$ is pure for all time $t$. Physically this means that if the initial state is pure and there is no information loss during the measurement process, then the state will always be pure.

Moreover, the evolution (\ref{conditionstate1}) preserves the Gaussian property. Thus  (\ref{conditionstate1}) can be fully depicted by the equations of the first- and second-moment  in the Heisenberg  picture whose dynamics are the same as    (\ref{conditionexpectation}) and (\ref{conditioncovariance}), while the initial values should correspond to the pure state $\hat{\rho}_0$, respectively.
Recall that the purity of the Gaussian state only depends on the second moment. Hence,  from the correspondence between the Schr$\ddot{\textrm{o}}$dinger picture and the Heisenberg picture for Gaussian states, if  (\ref{conditioncovariance}) has a unique stabilizing steady solution, it must correspond to a pure Gaussian state. Thus, from Theorem 2, if $[C,~A]$ is detectable, then the Riccati equation \eqref{conditioncovariance} has a unique stabilizing steady solution $V$, which can be considered as a conditional covariance matrix of some pure Gaussian state.
~~~$\blacksquare$

To prepare a Gaussian state, the approach via continuous measurement has the following advantages:

(i) From Theorem 3, to prepare a pure Gaussian state, the conditions that need to be satisfied via the continuous measurement are much weaker as compared with conditions given by the dissipation-induced approach.

(ii) For the unconditional approach, if $V_s$ is a unique steady pure state of  (\ref{uncondition}), then it must also be a unique steady pure state of the conditional covariance matrix equation (\ref{conditioncovariance}). This is because that $V_s$ being a unique steady pure state of  (\ref{uncondition}) is equivalent to that  (\ref{un1}) holds. This further yields $$V_s C^\top+M=0.$$ Thus,    (\ref{conditioncovariance}) and (\ref{uncondition}) have the same steady pure state.

(iii) Under the same system Hamiltonian and coupling operators, the conditional covariance matrix $V_{\text{c}}$ obtained via continuous measurement approach evolving under (\ref{conditioncovariance}) takes the form \cite{Giovanni2021} $$V_{\text{unc}}=V_c+\Sigma,$$ where $V_{\text{unc}}$ obeys  (\ref{uncondition}), and $$\Sigma=\textrm{E}[\bar{X}_c\bar{X}_c^\top+\bar{X}_c^\top\bar{X}_c]-\Big{(}\textrm{E}[\bar{X}_c]\textrm{E}[\bar{X}_c^\top]
      +\textrm{E}[\bar{X}_c^\top]\textrm{E}[\bar{X}_c]\Big{)},$$ with $\bar{X}_c$ obeying  (\ref{conditionexpectation}). Note that the Gaussian state which corresponds to $V_{\text{unc}}$ may be mixed.  Since most of the useful properties of a Gaussian state are determined by the covariance matrix,  generally the smaller the covariance matrix (in the matrix partial order) is, the more useful the Gaussian state is. Thus, continuous measurement approach can prepare more useful Gaussian states under the same system parameters.

To better illustrate the advantage of the conditional method in preparing pure Gaussian states, we give a single mode example.

\textit{\bf {Example 1.}} Consider the Hamiltonian as $$\hat{H}=\kappa\frac{\hat{a}^2+\hat{a}^{\dagger 2}+\hat{a}^\dagger \hat{a}+\hat{a}\hat{a}^\dagger}{2}=\kappa\hat{q}^2,$$ with $\kappa$ the damping rate of the cavity. The corresponding $G$ matrix is
$G=\begin{pmatrix}
     2\kappa & 0 \\
     0 & 0
\end{pmatrix}$.
It can be verified that there does not exist a  positive definite matrix $V$ such that it satisfies  the second equation of \eqref{un1}. In other words, if the system Hamiltonian is $\hat{H}=\kappa\hat{q}^2$, no matter how to choose the coupling operator $\hat{L}$, no pure Gaussian states can be prepared by utilizing the dissipation-induced (unconditional) approach.

Now we employ the conditional approach, and verify that pure Gaussian states can be prepared via continuous measurement. We take  $V=\frac{1}{2}\begin{pmatrix}
     1 & 0 \\
     0 & 1
\end{pmatrix}$ as an example. It is clear that $V$ corresponds to a pure Gaussian state. To prepare it, select the measurement operator as  $$\hat{L}=\sqrt{\kappa}\Big{(}(\sqrt{2}-\frac{\sqrt{2}}{2}i)\hat{a}-\frac{\sqrt{2}}{2}i\hat{a}^\dagger\Big{)}=\sqrt{\kappa}\Big{(}(1-i)\hat{q}
+i\hat{p}\Big{)}.$$  The corresponding parameters are $$\Lambda=\sqrt{\kappa}(1-i,~i),~ C=\sqrt{\kappa}(2,0),~  A-MC=\begin{pmatrix}
     \kappa & 0 \\
     0 & -\kappa
\end{pmatrix}.$$ It can be verified that $[C,~A-MC]$ is detectable, and  $V=\frac{1}{2}\begin{pmatrix}
     1 & 0 \\
     0 & 1
\end{pmatrix}$ is a solution of  (\ref{riccati}).

\subsection{Prepare a Gaussian state with desired covariance matrix}
In this subsection, we give an algebraic characterization of how to prepare a Gaussian state with the desired covariance matrix. Let the complex matrix $\Lambda$ be decomposed into
$$\Lambda=\textrm{R}+i \textrm{Im}$$
with its real part $\textrm{R}\in \mathbb{R}^{m\times2m}$ and imaginary part $\textrm{Im}\in \mathbb{R}^{m\times2m}$, respectively.

\begin{thm}
Let $V_s$ be a covariance matrix corresponding to a pure Gaussian state. Then, this is a unique stabilizing steady solution of   (\ref{conditioncovariance}) if the following two conditions hold: \\
(i)~$\text{Rank}\begin{pmatrix}
     \textrm{R}V_sJ  \\
   \textrm{  R}
\end{pmatrix}=2m$;\\
(ii)~$G=-\textrm{R}^\top \textrm{Im }-\textrm{Im}^\top \textrm{R} +2J^\top V_s \textrm{R}^\top \textrm{R} +2\textrm{R}^\top \textrm{R} V_s J$.\\
\end{thm}
{\bf  Proof:} We first prove that  $V_s$ is a solution of  (\ref{riccati}) if condition (ii) is satisfied. By substituting $\Lambda=\textrm{R}+i \textrm{Im}$ into \eqref{riccati}, this is equivalent to prove that $V_s$ satisfies
\begin{equation}\nonumber
\begin{aligned}
&J(G+\textrm{R}^\top \textrm{Im} +\textrm{Im}^\top \textrm{R})V_s+V_s(G+\textrm{R}^\top \textrm{Im} +\textrm{Im}^\top \textrm{R})J^\top\\
&-4V_s\textrm{R}^\top \textrm{R}V_s+J\textrm{R}^\top \textrm{R}J^\top=0.
\end{aligned}
\end{equation}
Multiplying the above equation by $J^\top$ from left and by $J$ from right, we have the following equivalent equation
\begin{equation}\label{JriccatiJ}
\begin{aligned}
&(G+\textrm{R}^\top\textrm{ Im} +\textrm{Im}^\top \textrm{R})V_sJ+J^\top V_s(G+\textrm{R}^\top \textrm{Im} +\textrm{Im}^\top \textrm{R})\\
&-4J^\top V_s\textrm{R}^\top \textrm{R}V_sJ+\textrm{R}^\top \textrm{R}=0.
\end{aligned}
\end{equation}
Since $V_s$ corresponds to a pure Gaussian state, we have \cite{Wolf2004} $$JV_sJV_s=-\frac{I}{4}.$$
Then $\textrm{R}^\top \textrm{R}$ can be expressed as
$$\textrm{R}^\top \textrm{R}=-2\textrm{R}^\top \textrm{R} V_sJV_sJ-2J^\top V_s J^\top V_s\textrm{R}^\top \textrm{R}.$$
Substituting this expression of $\textrm{R}^\top \textrm{R}$ into the left hand side of  (\ref{JriccatiJ}), we have
\begin{equation}\nonumber
\begin{aligned}
&~~~(G+\textrm{R}^\top\textrm{ Im} +\textrm{Im}^\top \textrm{R}-2J^\top V_s\textrm{R}^\top \textrm{R}-2\textrm{R}^\top \textrm{R} V_sJ)V_sJ\\
&+J^\top V_s(G+\textrm{R}^\top \textrm{Im }+\textrm{Im}^\top \textrm{R}-2J^\top V_s\textrm{R}^\top \textrm{R}-2\textrm{R}^\top \textrm{R} V_sJ)\\
&=0,
\end{aligned}
\end{equation}
if condition (ii) is met. Thus $V_s$ is a solution of  (\ref{riccati}).

Now we prove that $V_s$ is a  unique stabilizing steady solution of   (\ref{conditioncovariance}).  According to Theorem 2, we just need to prove that conditions (i)  and (ii) imply that $[C,~A-MC]$ is detectable. On the contrary, there exists  $\lambda$ with $\textrm{Re}(\lambda)\geq0$, and a corresponding eigenvector $x$, such that
\begin{equation}\nonumber
\begin{aligned}
(A-MC)x&=\lambda x,\\
\textrm{R}~x&=0.
\end{aligned}
\end{equation}
By utilizing condition (ii) and after some calculations, we have
\begin{equation}\nonumber
\begin{aligned}
2\textrm{R}^\top \textrm{R} V_s Jx=-\lambda Jx.
\end{aligned}
\end{equation}
Multiplying this equation by $x^\dagger J^\top V_s$ from left yields
\begin{equation}\label{counter1}
\begin{aligned}
2x^\dagger J^\top V_s\textrm{R}^\top \textrm{R }V_s Jx=-\lambda x^\dagger J^\top V_sJ x.
\end{aligned}
\end{equation}
Note that condition (i) $\text{Rank}\begin{pmatrix}
     \textrm{R}V_sJ  \\
     \textrm{R}
\end{pmatrix}=2m$ and $\textrm{R}x=0$ imply that $\textrm{R}V_sJx\neq0$. Thus, $x^\dagger J^\top V_s\textrm{R}^\top \textrm{R} V_s Jx>0$. In addition, since $V_s$ corresponds to a Gaussian state, this yields $V_s>0$, and  $x^\dagger J^\top V_sJ x>0$ accordingly.
Then from  (\ref{counter1}),  $\lambda<0$. But this contradicts with $\textrm{Re}(\lambda)\geq0$. Therefore, $[C,~A-MC]$ is detectable. Hence, $V_s$ is a unique stabilizing steady solution of   (\ref{conditioncovariance}) under conditions (i) and (ii).~~~$\blacksquare$
\begin{remark}
Under condition (ii) in Theorem 4, condition (i) holds if and only if $[C,~A-MC]$ is detectable. The necessity part has been proven in the proof of Theorem 4. To prove the sufficiency, we first note that under condition (ii), $[C,~A-MC]$ being detectable is equivalent to $[\textrm{R},~ J\textrm{R}^\top \textrm{R}V_sJ]$ being detectable. Further, if condition (i) does not hold, then there exists  $x\neq0$, such that $\textrm{R}V_sJx=0$ and $\textrm{R}x=0$. This clearly contradicts with $[\textrm{R},~ J\textrm{R}^\top \textrm{R}V_sJ]$ being detectable.
\end{remark}
\begin{remark} To make condition (i) of Theorem 4  hold, there are a lot of degrees of freedom to choose $\textrm{R}$. Here we give a simple choice such that  $\textrm{R}=\begin{pmatrix}
     I & 0
\end{pmatrix}$, where $I$ is the $m\times m$ identity matrix and $0$ is the $m\times m$ null matrix. Now let us check conditon (i). Given a target covariance matrix $V_s$ described by
$$
V_s=\begin{pmatrix}
     V_{11} & V_{12} \\
     V_{12}^\top & V_{22}
\end{pmatrix},
$$  since it corresponds to a pure Gaussian state,
$$V_{11}=V_{11}^\top>0~and~V_{22}=V_{22}^\top>0.$$ Then we have
$
\textrm{R}V_sJ=\begin{pmatrix}
     -V_{12} & V_{11}
\end{pmatrix}.
$
It is clear that
$$
\text{Rank}\begin{pmatrix}
     \textrm{R}V_sJ  \\
   \textrm{  R}
\end{pmatrix}=\textrm{Rank}\begin{pmatrix}
     -V_{12} & V_{11}\\
       I  & 0
\end{pmatrix}=2m.
$$
\end{remark}

\begin{remark}
Note that in Theorem 4, to prepare a desired pure Gaussian state with covariance matrix $V_s$, there is no restriction on the imaginary part of the coupling matrix $\Lambda$, which is helpful to relax the restrictions of  experimental realizations.
\end{remark}

\section{Imperfect detection efficiency}

In practical applications, we may not be able to achieve perfect detection efficiency and the imperfect efficiency $\eta<1$ may have significant impact on the performance in preparing Gaussian states. In this section we consider the case where the detection efficiency $\eta<1$. Here we assume that all the detection efficiencies are the same for simplicity.  In this case the dynamics of the conditional state $\hat{\rho}^c_t$ is described by
\begin{equation}\label{conditionstate2}
\begin{aligned}
d\hat{\rho}_t^c=&\mathcal{L}^\dagger[\hat{\rho}^c_t]dt\\+&\sqrt{\eta}\sum_{i=1}^m\Big{(}\hat{L}_i\hat{\rho}^c_t +\hat{\rho}^c_t \hat{L}^\dagger_i-\text{Tr}[\hat{\rho}^c_t (\hat{L}_i+\hat{L}_i^\dagger)]\hat{\rho}^c_t \Big{)}dw_{i,t},
\end{aligned}
\end{equation}
where the innovations $$dw_{i,t}=d{y}_{i,t}-\sqrt{\eta}\langle\hat{\rho}^c_t ,\hat{L}_i+\hat{L}_i^\dagger\rangle dt$$ are Gaussian, and
$$\mathcal{{L}}^\dagger[\hat{\rho}_t^c]=-i[\hat{H},\hat{\rho}_t^c]+\sum_{i=1}^{m}(\hat{L}_i
 \hat{\rho}_t^c \hat{L}_i^\dagger-\frac{1}{2}\hat{L}_i^\dagger \hat{L}_i \hat{\rho}_t^c-\frac{1}{2}\hat{\rho}_t^c\hat{L}_i^\dagger \hat{L}_i ).$$

If the initial state is pure Gaussian,  (\ref{conditionstate2}) still preserves the Gaussian nature. However, due to the imperfect detection, there is information loss during the measurement. This leads to that the Gaussian state  may become mixed as the state evolving.

For the $m$ mode system considered in subsection 2.3, the first- and second-moment equations corresponding to (\ref{conditionexpectation}) and (\ref{conditioncovariance}) become
\begin{equation}\label{covariance-eta}
\begin{aligned}
d\bar{X}_t&=(A\bar{X}_t+Bu_t)dt+\sqrt{\eta}(V_t C^\top+M)d\tilde{y}_t,\\
\frac{d}{dt}V_t&=AV_t+V_tA^\top+N-\eta (V_t C^\top+M)(V_t C^\top+M)^\top.
\end{aligned}
\end{equation}
If $V$ is a stabilizing steady solution of (\ref{covariance-eta}), then $V$ needs to satisfy the following Riccati equation
\begin{equation}\nonumber
AV+VA^\top+N-\eta(V C^\top+M)(V C^\top+M)^\top=0,
\end{equation}
which is equivalent to
\begin{equation}\label{riccati-eta}
\begin{aligned}
(A-\eta MC)V&+V(A-\eta MC)^\top\\&-\eta VC^\top CV+N-\eta MM^\top =0.
\end{aligned}
\end{equation}
Similar to Theorem 2 where the detection efficiency $\eta=1$, we have the following conclusion.
\begin{thm}
The Riccati equation \eqref{covariance-eta} has a unique stabilizing steady solution satisfying $V\geq\pm\frac{i}{2}J$ if and only if $[C,~A-MC]$ is detectable.
\end{thm}

To prove Theorem 5, we first rearrange \eqref{covariance-eta} into the following form
\begin{equation}\nonumber
(A-\eta MC)V+V(A-\eta MC)^\top-\eta VC^\top CV+Q_V^\top Q_V=0,
\end{equation}
where $Q_V=\begin{bmatrix}
     \textrm{R}  \\
     \sqrt{1-\eta}~\textrm{Im}
\end{bmatrix}J^\top$. We have the following lemma.

\begin{lemma}\label{lemma5}
$[Q_V,~(A-\eta MC)^\top]$ has no unobservable modes on the imaginary axis if $[C,~A-MC]$ is detectable.
\end{lemma}

{\bf  Proof:} Suppose, on the contrary, that  there exists $\lambda\in \mathbb{R}$ and a corresponding vector $x\neq 0$, such that
\begin{equation}\nonumber
\begin{aligned}
&(A-\eta MC)^\top x=i \lambda x,\\
&Q_V x=\begin{bmatrix}
     \textrm{R}  \\
     \sqrt{1-\eta}\textrm{Im}
\end{bmatrix}J^\top x=0.
\end{aligned}
\end{equation}
We have $CJ^\top x=2RJ^\top x=0$. Moreover, since $$(A-\eta MC)J^\top=J(A-\eta MC)^\top,$$ we have
\begin{equation}\nonumber
(A-\eta MC)J^\top x=J(A-\eta MC)^\top x=i \lambda J x,
\end{equation}
which yields $$(A-\eta MC)J^\top x=-i \lambda J^\top x.$$
Combining this conclusion with $CJ^\top x=0$ contradicts with the condition that $[C,~A-MC]$ is detectable.
Thus, $[Q_V,~(A-\eta MC)^\top]$ has no unobservable modes on the imaginary axis if $[C,~A-MC]$ is detectable.~~~$\blacksquare$

Secondly, similar to  \eqref{riccatiY}, denoting  $Y=V-\frac{i}{2}J$,  we have the Riccati equation about $Y$ as
\begin{equation*}\label{riccati-eta-Y}
\begin{aligned}
&(A-\eta MC-\frac{i}{2}\eta J C^\top C)Y+Y(A-\eta MC-\frac{i}{2}\eta J C^\top C)^\top\\
&-\eta YC^\top CY+Q_Y^\top Q_Y=0,
\end{aligned}
\end{equation*}
where $Q_Y=\sqrt{1-\eta}\begin{bmatrix}
     \textrm{R}  \\
     \textrm{Im}
\end{bmatrix}J^\top$. Then we have the following lemma.
\begin{lemma}\label{lemma6}
$[Q_V,~(A-\eta MC)^\top]$ has no unobservable modes on the imaginary axis if and only if  $[Q_Y,~(A-\eta MC-\frac{i}{2}\eta J C^\top C)^\top]$ has no unobservable modes on the imaginary axis.
\end{lemma}

{\bf  Proof:}
\textit{Sufficiency:} Suppose, on the contrary, that there exists $\lambda\in \mathbb{R}$ and a corresponding vector $x\neq 0$, such that
\begin{equation}\nonumber
\begin{aligned}
&(A-\eta MC)^\top x=i \lambda x,\\
&Q_V x=\begin{bmatrix}
     \textrm{R}  \\
     \sqrt{1-\eta}\textrm{Im}
\end{bmatrix}J^\top x=0.
\end{aligned}
\end{equation}
Then we have $CJ^\top x=2\textrm{R}J^\top x=0$. Therefore,
\begin{equation}\nonumber
\begin{aligned}
(A-\eta MC-\frac{i}{2}\eta J C^\top C)^\top x&=(A-\eta MC)^\top x
=i \lambda x,\\
Q_Y x&=0,
\end{aligned}
\end{equation}
which contradicts with the condition that $[Q_Y,(A-\eta MC-\frac{i}{2}\eta J C^\top C)^\top]$ has no unobservable modes on the imaginary axis.

\textit{Necessity:} Suppose there exists $\lambda\in \mathbb{R}$ and a corresponding vector $x\neq 0$, such that
\begin{equation}\nonumber
\begin{aligned}
&(A-\eta MC-\frac{i}{2}\eta J C^\top C)^\top x=i \lambda x,\\
&Q_Y x=\sqrt{1-\eta}\begin{bmatrix}
     \textrm{R}  \\
    \textrm{ Im}
\end{bmatrix}J^\top x=0.
\end{aligned}
\end{equation}
Then we have $CJ^\top x=0$. Therefore,
\begin{equation}\nonumber
\begin{aligned}
(A-\eta MC)^\top x&=(A-\eta MC-\frac{i}{2}\eta J C^\top C)^\top x
=i \lambda x,\\
Q_V x&=0,
\end{aligned}
\end{equation}
which  contradicts with the condition that  $[Q_V,~(A-\eta MC)^\top]$ has no unobservable modes on the imaginary axis.

Thus, $[Q_V,~(A-\eta MC)^\top]$ has no unobservable modes on the imaginary axis if and only if  $[Q_Y,~(A-\eta MC-\frac{i}{2}\eta J C^\top C)^\top]$ has no unobservable modes on the imaginary axis.~~~$\blacksquare$

Now we give the  proof of  Theorem 5.

{\bf  Proof:} \textit{Sufficiency:} If $[C,~A-MC]$ is detectable, then from Lemma \ref{lemma5}, $[Q_V,~(A-\eta MC)^\top]$ has no unobservable modes on the imaginary axis. Thus according to Theorem 13.7 in \cite{zhou2002}, the Riccati equation \eqref{riccati-eta} about $V$ has a unique stabilizing solution which is positive semidefinite.

To prove $V\geq\frac{i}{2}J$, from Lemma \ref{lemma3}, we just need to verify that $[C,~A-MC-\frac{i}{2}\eta J C^\top C]$ is detectable and $[Q_Y,(A-\eta MC-\frac{i}{2}\eta J C^\top C)^\top]$ has no unobservable modes on the imaginary axis. In fact, as $[C,~A-MC]$ is detectable, so is $[C,~A-MC-\frac{i}{2}\eta J C^\top C]$. Moreover, from Lemmas \ref{lemma5} and \ref{lemma6},  $[Q_Y,(A-\eta MC-\frac{i}{2}\eta J C^\top C)^\top]$ has no unobservable modes on the imaginary axis.  Thus,  the Riccati equation \eqref{covariance-eta}  has a unique stabilizing steady solution satisfying $V\geq\pm\frac{i}{2}J$ if  $[C,~A-MC]$ is detectable.

\textit{Necessity:} According to \cite{zhou2002}, the solution of \eqref{riccati-eta} has the following form
\begin{equation}\nonumber
\begin{aligned}
Y=\int_0^{\infty}&\textrm{e}^{\big{(}(A-\eta MC)^\top -\eta C^\top CV\big{)}^\top t}\big{(}\eta VC^\top CV\\
&+N-\eta MM^\top\big{)} \textrm{e}^{\big{(}(A-\eta MC)^\top -\eta C^\top CV\big{)}t}.
\end{aligned}
\end{equation}
It is clear that $(A-\eta MC)^\top -\eta C^\top CV$ is stable. This yields $[C,~A-MC]$ being detectable.~~~$\blacksquare$

Although the Gaussian state may be not pure if the detection is imperfect, the steady solution $V_{\text{c}}$ of the conditional covariance matrix  equation (\ref{covariance-eta}) is still smaller than the steady solution $V_{\text{unc}}$  obtained via the unconditional dissipation-induced approach (\ref{uncondition}).

\section{Conclusion}
In this paper we consider the problem of how to prepare Gaussian states via continuous measurement. We present some necessary and sufficient conditions for the conditional dynamics of the covariance matrix to have a unique stabilizing steady solution. This conditional method is much superior to the unconditional approaches in  the following two aspects. On one hand, the conditions given by our method is much weaker than those given by the unconditional methods, which may be beneficial to experimental realizations. On the other hand,  under the same system parameters, the Gaussian states prepared via the continuous measurement may be more useful in quantum information processing.

\begin{ack}                               % Place acknowledgements
This work was supported by the National Natural Science Foundation of China (Nos.  11688101, 61773370, 61833010 and 61621003) and the Australian Research Councils Discovery Projects funding scheme under Project DP190101566.
\end{ack}

\appendix

\section*{Appendix}

\section{Riccati equation in the complex domain}
In this Appendix, we present some results concerning a matrix Ricaati equation in the complex domain, which are similar to those in Section 13.2 of \cite{zhou2002}, where the domain considered is real.

Let $F\in \mathbb{C}^{n\times n},~P=P^\dagger\in \mathbb{C}^{n\times n}$ and $K=K^\dagger\in\mathbb{C}^{n\times n}$. The matrix Riccati equation to be considered is
\begin{equation}\label{r1}
F^\dagger X+XF+XP X+K=0.
\end{equation}
We associate a $2n\times 2n$ matrix in the complex domain with the Riccati equation (\ref{r1}) as
\begin{equation}\nonumber
H=\begin{bmatrix}
     F & P \\
     -K & -F^\dagger
\end{bmatrix}.
\end{equation}
Noting that the $2n\times 2n$ matrix
\begin{equation}\nonumber
J=\begin{bmatrix}
     0 & I \\
     -I & 0
\end{bmatrix}
\end{equation}
has the property $J^2=-I$. Then we have $$J^{-1}HJ=-JHJ=-H^\dagger.$$ Thus, $H$ and $-H^\dagger$ are similar. This  further implies that $\lambda$ is an eigenvalue of $H$ if and only if  $-{\lambda}^*$ is, where $\lambda^*$ denotes the conjugate of $\lambda$.

Assume that $H$ has no eigenvalues on the imaginary axis. Then it must have $n$ eigenvalues in the left half plane and $n$ eigenvalues in the right half plane.  Denote the corresponding two $n$-dimensional spectral subspaces as $\mathcal {X}_{-}(H)$ and $\mathcal {X}_{+}(H)$, respectively. To be specific, the former is the invariant subspace corresponding to eigenvalues in the left half plane and the latter corresponds to eigenvalues in the right half plane. By finding a basis of $\mathcal {X}_{-}(H)$, stacking the basis vectors up to form a matrix, and partitioning the matrix, we have
\begin{equation}\nonumber
\mathcal{X}_{-}(H)=\textit{Span}
\begin{bmatrix}
     X_1  \\
     X_2
\end{bmatrix},
\end{equation}
where $X_1,~X_2 \in \mathbb{C}^{n\times n}$. If $X_1$ is nonsingular,
% or, equivalently, if the two subspaces
%\begin{equation}\label{twosubspaces}
%\mathcal{X}_{-}(H),~\textit{Im}\begin{bmatrix}
%    0  \\
%     I
%\end{bmatrix}
%\end{equation}
%are complementary,
we can define $$X=X_2X_1^{-1}.$$ It can be verified that $X$ is independent of a specific choice of basis of $\mathcal{X}_{-}(H)$, and is \textit{uniquely} determined by $H$,  which can be described by the map $$\textit{Ric}: H\rightarrow X.$$
We will take the domain of \textit{Ric}, denoted by $\textrm{dom}(\textit{Ric})$, to consist of matrices $H$ with the following two properties:
\begin{itemize}
  \item \textit{Stability property}: $H$ has no eigenvalues on the imaginary axis;
  \item \textit{Complementarity}: the two subspaces  $\mathcal {X}_{-}(H)$ and $\textit{span}\begin{bmatrix}
    0  \\
     I
\end{bmatrix}$ are complementary, which is equivalent to $X_1$ being nonsingular.
\end{itemize}
Now we have Theorem 6 concerning the properties of $X$.

\begin{thm}\label{thm13.5}
Suppose $H\in \textrm{dom}(\textit{Ric})$ and $X=\textit{Ric}(H)$. Then

(i) $X=X^\dagger$;

(ii)  $X$ satisfies Riccati equation (\ref{r1}), i.e.,
\begin{equation}\nonumber
F^\dagger X+XF+XP X+K=0,
\end{equation}

(iii) $F+PX$ is stable.
\end{thm}

{\bf  Proof:}
(i) Since $H\in \textrm{dom}(\textit{Ric})$, there exists a stable matrix denoted by $H_{-} \in \mathbb{C}^{n\times n}$, which is a matrix representation of $H|_{\mathcal {X}_{-}(H)}$, such that
\begin{equation}\label{HX1X2}
H\begin{bmatrix}
    X_1  \\
     X_2
\end{bmatrix}=\begin{bmatrix}
    X_1  \\
     X_2
\end{bmatrix}H_{-}.
\end{equation}
Multiplying this equation from left by $\begin{bmatrix}
    X_1  \\
     X_2
\end{bmatrix}^\dagger J$ yields

\begin{equation}\nonumber
\begin{bmatrix}
    X_1  \\
     X_2
\end{bmatrix}^\dagger JH\begin{bmatrix}
    X_1  \\
     X_2
\end{bmatrix}=\begin{bmatrix}
    X_1  \\
     X_2
\end{bmatrix}^\dagger J\begin{bmatrix}
    X_1  \\
     X_2
\end{bmatrix}H_{-}.
\end{equation}
Noting that $JH=(JH)^\dagger$, the left hand side of the above equation is Hermitian, and so is the right hand side. Thus, we have
\begin{equation}\nonumber
\begin{aligned}
(-X_1^\dagger X_2+X_2^\dagger X_1)H_{-}&=H_{-}^\dagger (-X_1^\dagger X_2+X_2^\dagger X_1)^\dagger\\
&=-H_{-}^\dagger (-X_1^\dagger X_2+X_2^\dagger X_1).
\end{aligned}
\end{equation}
This is a Lyapunov equation. Since $H_{-}$ is stable, the unique solution is
\begin{equation}\nonumber
-X_1^\dagger X_2+X_2^\dagger X_1=0.
\end{equation}
Since $X_1$ is nonsingular, $X$ can be represented as $$X=(X_1^\dagger)^{-1}X_1^\dagger X_2X_1^{-1}.$$ This combined with $X_1^\dagger X_2=X_2^\dagger X_1$ yields $X=X^\dagger$.

(ii) Start with the equation
\begin{equation}\nonumber
H\begin{bmatrix}
    X_1  \\
     X_2
\end{bmatrix}=\begin{bmatrix}
    X_1  \\
     X_2
\end{bmatrix}H_{-}.
\end{equation}
Multiplying the above equation from right by $X_1^{-1}$ yields
\begin{equation}\label{HIX}
H\begin{bmatrix}
    I  \\
    X
\end{bmatrix}=\begin{bmatrix}
    I  \\
     X
\end{bmatrix}X_1 H_{-} X_1^{-1}.
\end{equation}
Now multiplying from left by $\begin{bmatrix}
    X  &  -I
\end{bmatrix}$, we have
\begin{equation}\nonumber
\begin{bmatrix}
    X  &  -I
\end{bmatrix} H \begin{bmatrix}
    I  \\
    X
\end{bmatrix}=0.
\end{equation}
This is precisely the Riccati equation (\ref{r1}).

(iii) Multiplying \eqref{HIX}  from left by $\begin{bmatrix}
    I  &  0
\end{bmatrix}$, we have
\begin{equation}\nonumber
F+PX=X_1H_{-}X_1^{-1}.
\end{equation}
Thus $F+PX$ is stable because $H_{-}$ is stable.~~~$\blacksquare$

A solution $X$ of (\ref{r1}) is called \textit{stabilizing} if $F+PX$ is stable. A necessary and sufficient condition for the existence of a unique stabilizing solution of  (\ref{r1}) is stated in the following theorem.

\begin{thm}\label{thm13.6}
Suppose $H$ has no imaginary eigenvalues and $P$ is either positive semi-definite or negative semi-definite. Then $H \in \textrm{dom}(\textit{Ric})$ if and only if $[P,~F^\dagger]$ is detectable.
\end{thm}

{\bf  Proof:}
\textit{Sufficiency:} To prove that $H \in \textrm{dom}(\textit{Ric})$, we just need to show that
$X_1$ is nonsingular, i.e., $\textrm{Ker}(X_1) = 0$.

First, let us demonstrate that $\textrm{Ker}( X_1)$ is $H_{-}$-invariant, where $H_{-}$ is given in (\ref{HX1X2}). To prove this, let $x \in \textrm{Ker}(X_1)$. Multiplying \eqref{HX1X2} from left by $\begin{bmatrix}
    I  &  0
\end{bmatrix}$ yields
\begin{equation}\label{AX1+PX2}
FX_1 + PX_2 = X_1H_{-}.
\end{equation}
Multiplying the above equation from left by $x^\dagger X_2^\dagger$, and by $x$ from right, and recalling the fact that $X_2^\dagger X_1=(X_2^\dagger X_1)^\dagger$, we have
\begin{equation}\nonumber
x^\dagger X_2^\dagger P X_2 x=0.
\end{equation}
Since $P$ is semi-definite, this implies that $P X_2 x=0$. Then, from \eqref{AX1+PX2}, we have $X_1 H_{-}x=0$, i.e., $H_{-}x\in \textrm{Ker}(X_1)$, which proves that $\textrm{Ker}(X_1)$ is $H_{-}$-invariant.

Now to prove that $X_1$ is nonsingular, suppose, on the contrary, that $\textrm{Ker}(X_1) \neq 0$. Then $H_{-}|_{\textrm{Ker}(X_1)}$ has an eigenvalue $\lambda$ with $\textrm{Re}(\lambda)<0$, and a corresponding eigenvector $x\neq 0$, i.e.,
\begin{equation}\nonumber
\begin{aligned}
&H_{-}x=\lambda x, ~\textrm{Re} (\lambda)<0,\\
& x\in \textrm{Ker}(X_1),~x \neq0.
\end{aligned}
\end{equation}
Multiplying \eqref{HX1X2} from left by $\begin{bmatrix}
    0  &  I
\end{bmatrix}$ yields
\begin{equation}\nonumber
-K X_1-F^\dagger X_2=X_2H_{-}.
\end{equation}
Multiplying the above equation by $x$ from right, we have
\begin{equation}\nonumber
(F^\dagger+\lambda I)X_2x=0.
\end{equation}
Since $PX_2 x=0$, we have
\begin{equation}\nonumber
\begin{bmatrix}
    F^\dagger+\lambda I  \\
    P
\end{bmatrix}X_2x=0.
\end{equation}
Then the detectability of $[P,~F^\dagger]$ implies $X_2x=0$, which contradicts with the fact that $\begin{bmatrix}
    X_1  \\
     X_2
\end{bmatrix}$ has full column rank.

\textit{Necessity:} From Theorem 6,  $H\in \textrm{dom}(\textit{Ric})$ implies that $X$ is a stabilizing solution such that $F+PX$ is stable. From (v) of Lemma 1, this further implies that $[P,~F^\dagger]$ is detectable.~~~$\blacksquare$

\begin{thm}
Suppose $H$ has the form
\begin{equation}\nonumber
H=\begin{bmatrix}
    F & -OO^\dagger  \\
    -Z^\dagger Z & -F^\dagger
\end{bmatrix}.
\end{equation}
Then,

(i) $H\in \textrm{dom}(\textit{Ric})$ if and only if $[O^\dagger,~F^\dagger]$ is detectable and $[Z,~F]$ has no unobservable modes on the imaginary axis;

(ii) $X=\textit{Ric}(H)\geq0$ if $H\in \textrm{dom}(\textit{Ric})$;

(iii) $H\in \textrm{dom}(\textit{Ric})$, then  $\textrm{Ker}(X)=0$ if and only if $[Z,~F]$ has no stable unobservable modes.

\end{thm}

{\bf  Proof:} To prove (i), noting that from Theorem \ref{thm13.6},  the detectability of $[O^\dagger,~F^\dagger]$ is necessary. Hence, from Theorem 7, we only need to show that under the condition $[O^\dagger,~F^\dagger]$ being detectable, $H$ has no imaginary eigenvalues if and only if $[Z,~F]$ has no unobservable modes on the imaginary axis.

Suppose that there exists a real $\omega$ such that $i \omega$ is an eigenvalue of $H$, and $\begin{bmatrix}
    x  \\
    z
\end{bmatrix}\neq0$ is the corresponding eigenvector. Then, we have
\begin{equation}\nonumber
\begin{aligned}
Fx-OO^\dagger z=i \omega x,\\
-Z^\dagger Zx-F^\dagger z=i\omega z.
\end{aligned}
\end{equation}
Re-arranging the above equation yields
\begin{equation}\label{OzZx}
\begin{aligned}
(F-i\omega I)x=OO^\dagger z,\\
-(F-i\omega I)^\dagger z=Z^\dagger Zx.
\end{aligned}
\end{equation}
Thus
\begin{equation}\nonumber
\begin{aligned}
z^\dagger(F-i\omega I)x= z^\dagger OO^\dagger z=||O^\dagger z||^2,\\
-x^\dagger (F-i\omega I)^\dagger z= x^\dagger Z^\dagger Zx=||Zx||^2.
\end{aligned}
\end{equation}
Hence, $\ x^\dagger(F-i\omega I)^\dagger z$ is real and
\begin{equation}\nonumber
-||Zx||^2=x^\dagger(F-i\omega I)^\dagger z=z^\dagger (F-i\omega I) x=||O^\dagger z||^2.
\end{equation}
Therefore, $Zx=0$ and $O^\dagger z=0$. From \eqref{OzZx},
\begin{equation}\nonumber
\begin{aligned}
(F-i\omega I)x=0,\\
-(F-i\omega I)^\dagger z=0.
\end{aligned}
\end{equation}
Combining this with  $Zx=0$ and $O^\dagger z=0$, we have
\begin{equation}\nonumber
\begin{aligned}
\begin{bmatrix}
    F-i\omega I  \\
    Z
\end{bmatrix}x=0,\\
\begin{bmatrix}
    F^\dagger+i\omega I  \\
    O^\dagger
\end{bmatrix}z=0.
\end{aligned}
\end{equation}
Under the condition that $[O^\dagger,~F^\dagger]$ is detectable,  we have $z=0$, and  $x\neq 0$. Now it is straightforward to see that $i\omega$ is an eigenvalue of $H$ if and only if  $i\omega$ is an unobservable mode of $[Z,~F]$.

To prove (ii), let $X=\textit{Ric}(H)$.  The Riccati equation is
\begin{equation}\nonumber
F^\dagger X+XF-XOO^\dagger X+Z^\dagger Z=0,
\end{equation}
or equivalently,
\begin{equation}\label{ricctiFOZ}
(F-OO^\dagger X)^\dagger X+X(F-OO^\dagger X)+XOO^\dagger X+Z^\dagger Z=0.
\end{equation}
Noting that $F-OO^\dagger X$ is stable by  (iii) of Theorem \ref{thm13.5}, we have
\begin{equation}\nonumber
\begin{aligned}
X=&\int_0^{\infty}\textrm{e}^{(F-OO^\dagger X)^\dagger t}(XOO^\dagger X+Z^\dagger Z )\textrm{e}^{(F-OO^\dagger X)t}dt.
\end{aligned}
\end{equation}
Since $XOO^\dagger X+Z^\dagger Z $ is positive semi-definite, so is $X$.

Finally, we prove (iii).

\textit{Sufficiency:} Let us first show that  $\textrm{Ker}(X)$ is an $F$-invariant subspace.  Suppose $x\in \textrm{Ker} (X)$, then $Xx=0$. Multiplying \eqref{ricctiFOZ} from left by $x^\dagger$ and right by $x$ yields
\begin{equation}\nonumber
Zx=0.
\end{equation}
Now multiply \eqref{ricctiFOZ} from right by $x$ to get
\begin{equation}\nonumber
XFx=0.
\end{equation}
Thus, $\textrm{Ker}(X)$ is  $F$-invariant.

Now if $\textrm{Ker}(X)\neq 0$, then there exists a nonzero  $ x\in \textrm{Ker}(X)$ and a corresponding $\lambda$ such that
\begin{equation*}
\begin{aligned}
Fx&=(F-OO^\dagger X)x=\lambda x,\\
Zx&=0.
\end{aligned}
\end{equation*}
From (iii) of Theorem 6,  $F-OO^\dagger X$ is stable, so $\textrm{Re}(\lambda)<0$. Thus, $\lambda$ is a stable unobservable mode, which is a contradiction.

\textit{Necessity:} Suppose, on the contrary, that $[Z,~F]$ has an unobservable stable mode $\lambda$, i.e., there is $x\neq0$ such that $Fx=\lambda x,~\textrm{Re}(\lambda)<0,$ and $Zx=0$. By multiplying the Riccati equation \eqref{ricctiFOZ} from left by $x^\dagger$ and right by $x$, we have
\begin{equation}\nonumber
2\textrm{Re} (\lambda) x^\dagger Xx-x^\dagger X OO^\dagger Xx=0.
\end{equation}
Since $X$ is positive semi-definite, i.e., $x^\dagger Xx\geq0$, and $x^\dagger X OO^\dagger Xx \geq0$, we have $x^\dagger Xx=0$, i.e., $X$ is singular, which is a contradiction.~~~$\blacksquare$

\section{Proof of Lemma \ref{lemma3}}
It is straightforward to see that Lemma \ref{lemma3} can be obtained from (i), (ii) of Theorem 8 and (iii) of Theorem 6.

%\bibliographystyle{plain}        % Include this if you use bibtex
%\bibliography{123}           % and a bib file to produce the
                                 % bibliography (preferred). The
                                 % correct style is generated by
                                 % Elsevier at the time of printing.

%\begin{thebibliography}{99}     % Otherwise use the
                                 % thebibliography environment.
                                 % Insert the full references here.
                                 % See a recent issue of Automatica
                                 % for the style.

\end{document}